\documentclass[conference]{IEEEtran}
\IEEEoverridecommandlockouts
\usepackage{cite}
\usepackage{amsmath,amssymb,amsfonts}
\usepackage{algorithmic}
\usepackage{graphicx}
\usepackage{textcomp}
\usepackage{float}
\usepackage{xcolor}
\usepackage{url}
\usepackage{flushend}
\def\BibTeX{{\rm B\kern-.05em{\sc i\kern-.025em b}\kern-.08em
    T\kern-.1667em\lower.7ex\hbox{E}\kern-.125emX}}
\begin{document}

\title{LLM vs.\ Human Unit Tests: Fault Detection on Real Python Bugs}

\author{\IEEEauthorblockN{Phouvadeth Vathana}
\IEEEauthorblockA{\textit{University of Tennessee} \\
Knoxville, USA \\
pvathana@vols.utk.edu}
\and
\IEEEauthorblockN{Prapti Bhatt}
\IEEEauthorblockA{\textit{University of Tennessee} \\
Knoxville, USA \\
pbhatt1@vols.utk.edu}
\and
\IEEEauthorblockN{Rishi Patel}
\IEEEauthorblockA{\textit{University of Tennessee} \\
Knoxville, USA \\
rpatel92@vols.utk.edu}
\and
\IEEEauthorblockN{Nasir U. Eisty}
\IEEEauthorblockA{\textit{University of Tennessee} \\
Knoxville, USA \\
neisty@utk.edu}
}

\maketitle

\begin{abstract}
Large language models (LLMs) have shown considerable promise for automated unit test generation, yet their practical effectiveness relative to human-written tests remains poorly understood. Existing evaluations commonly rely on coverage-oriented benchmarks that do not assess fault-detection capability directly. We present an empirical comparison of LLM-generated and human-written unit tests across three complementary Python benchmarks: 29 real historical bugs from BugsInPy, a function-level benchmark drawn from \texttt{python-slugify} and \texttt{packaging}, and a controlled paired benchmark. Our generation pipeline couples Gemini~2.5~Flash with a lightweight lexical retrieval mechanism that supplies bug-relevant context at generation time. Across eight quality dimensions, LLM-generated tests with retrieval-augmented context detect faults in 69\% of cases compared to 17.2\% for general-purpose human-written tests (Fisher's exact, $p < 0.001$, Cohen's~$h = 1.10$). Critically, line and branch coverage are nearly identical between the two approaches (84.8\% vs.\ 88.5\% and 75.2\% vs.\ 82.1\%), confirming that coverage is an insufficient proxy for fault-detection capability. We discuss the conditions under which each approach excels, characterize their complementary strengths, and identify the critical role of retrieval context and reproducible benchmark construction in meaningful test-quality evaluation.
\end{abstract}

\section{Introduction}

Software testing is a cornerstone of software quality assurance, yet writing unit tests is among the most repetitive and time-consuming activities in a developer's workflow. Unit tests verify individual functions in isolation, document expected behavior, and form the first line of defense against regressions when code changes. Despite their importance, test suites in practice are frequently incomplete: developers work under schedule pressure, edge cases are overlooked, and tests targeting specific fault conditions are rarely written until after a bug is discovered. The cost of this gap is significant; defects that escape to production are orders of magnitude more expensive to resolve than those caught during development~\cite{b14}.

Large language models (LLMs) have emerged as a promising technology for automating unit test generation. Recent work demonstrates that models such as GPT-4 and Gemini can produce syntactically valid, compilable tests for real-world functions with little or no manual prompting~\cite{b8,b9,b4}. Commercial tools now embed LLM-based test generation directly into developer workflows, raising expectations that automated generation can meaningfully reduce the human effort required to maintain quality test suites. These developments make a rigorous, evidence-based understanding of LLM test quality both timely and important.

However, the dominant evaluation framework in this area is inadequate. Most existing studies measure structural coverage and compilability, which assess whether generated code executes without errors but not whether it actually detects faults~\cite{b9,b8}. Inozemtseva and Holmes~\cite{b17} showed empirically that coverage and test-suite effectiveness are not strongly correlated, and Cai and Lyu~\cite{b7} demonstrated that tests can achieve high line coverage while leaving critical fault conditions entirely untested. Despite these findings, coverage remains the primary yardstick in LLM test generation research, producing an evaluation literature that may significantly overestimate practical test quality. A test that executes 90\% of a function's lines but asserts nothing meaningful about its output provides little value to a developer trying to prevent regressions.

A second limitation of existing studies is that they rarely evaluate LLM-generated tests in the context where automated generation is most naturally valuable: regression testing at the time a bug is fixed. When a developer resolves a defect, they possess precisely the information that an LLM needs to generate a precise, fault-targeted test: the bug report, the patch diff, and the surrounding source context. Evaluating LLMs in this context, where bug-relevant information is retrieved and supplied at generation time through a retrieval-augmented generation (RAG) pipeline, is both more realistic and more informative than evaluating direct prompting in isolation. Yet few studies have systematically compared RAG-augmented LLM tests against existing human-written tests on real historical faults.

A third challenge is benchmark quality. Real software repositories are noisy and heterogeneous. Bugs must be selected carefully to ensure reproducibility, repository versions must be checked out correctly, and relevant code and test artifacts must be isolated before any fair comparison can take place. Studies that treat dataset construction as a preprocessing detail introduce inconsistencies that can confound results and prevent replication. We argue that benchmark construction is itself a methodological contribution, and we describe our pipeline in sufficient detail to support independent reproduction.

This paper addresses all three challenges. We design and execute a multi-benchmark empirical study that evaluates LLM-generated tests on their ability to detect real Python bugs, not merely to achieve coverage. Our generation pipeline couples Gemini~2.5~Flash~\cite{b20} with a lightweight lexical RAG mechanism that retrieves patch diffs, bug descriptions, and surrounding source context at generation time. We compare the resulting tests against existing general-purpose human-written tests across eight quality dimensions, and we situate our findings within a clear account of when LLM generation provides the greatest value.

Our results reveal a striking and consequential finding. LLM-generated tests with retrieval-augmented context detect faults in 69\% of 29 real historical bugs compared to only 17.2\% for human-written tests, a four-fold difference significant at $p < 0.001$ with a large effect size (Cohen's~$h = 1.10$). Yet line and branch coverage are statistically indistinguishable between the two approaches (84.8\% vs.\ 88.5\% and 75.2\% vs.\ 82.1\%). This combination of near-equal coverage and dramatically different fault detection provides the clearest empirical evidence to date that coverage metrics are an unreliable proxy for test quality, and it points to a concrete deployment strategy: LLM test generation is most valuable at fix time, when bug context is available.

\noindent\textbf{Research Questions.} Three questions guide the study:

\noindent\hangindent=1.2cm\hangafter=1
\textbf{RQ1}\hspace{0.5em} How effective are context-aware LLM-generated tests at detecting real historical bugs compared to general-purpose human-written tests?

\noindent\hangindent=1.2cm\hangafter=1
\textbf{RQ2}\hspace{0.5em} Under what conditions do LLM-generated tests outperform or trail human-written tests?

\noindent\hangindent=1.2cm\hangafter=1
\textbf{RQ3}\hspace{0.5em} How do LLM-generated and human-written tests differ in terms of coverage, assertion density, documentation, and testing patterns?

\noindent\textbf{Contributions.} This paper makes four contributions:
\begin{itemize}
    \item A multi-benchmark evaluation pipeline integrating 29 real historical bugs from BugsInPy, a function-level open-source benchmark, and a controlled paired benchmark, together with a replication package.
    \item Empirical evidence that LLM-generated tests with retrieval-augmented context achieve a four-fold higher fault-detection rate than general-purpose human tests, while showing near-parity on structural coverage, directly challenging coverage as the primary evaluation criterion for test quality.
    \item A characterization of the conditions under which LLM-generated and human-written tests each excel, providing practical guidance for teams considering LLM-based test generation in continuous integration workflows.
    \item An articulation of benchmark construction as a first-class methodological concern in test-quality studies, with concrete recommendations for reproducible benchmark design.
\end{itemize}

\section{Related Work}

\subsection{LLM-Based Test Generation}

Sch\"{a}fer et al.~\cite{b8} conducted a large empirical evaluation of LLM-based unit test generation across multiple languages and found that generated tests achieve competitive branch coverage but exhibit semantic gaps and test smells relative to developer-written tests. Yang et al.~\cite{b9} evaluated multiple LLMs on the EvoEval benchmark and reported that while pass rates and coverage are reasonable, LLMs struggle with deep behavioral assertions. Ou\'{e}draogo et al.~\cite{b4} studied 310 Python projects at scale and found substantial variation in generated test quality depending on function complexity and project type; their companion study~\cite{b10} catalogued recurring test smells in LLM output. A persistent limitation of these evaluations is their reliance on structural coverage as the primary quality criterion. Our work shifts the central criterion to fault detection on real historical bugs, following the empirical argument of Inozemtseva and Holmes~\cite{b17} that coverage and test-suite effectiveness are not strongly correlated. Haroon et al.~\cite{b5} recently studied LLM test generation under software evolution and found that generated tests degrade in fault-detection ability as code changes; our study complements this by examining the baseline comparison on fixed historical defects.

\subsection{Traditional Automated Test Generation}

Before LLMs, search-based tools such as EvoSuite~\cite{b19} demonstrated that automated approaches can achieve high structural coverage for Java. These tools optimize primarily for coverage and line execution, however, and tend to produce tests with weak or absent semantic assertions~\cite{b11}. Watson et al.~\cite{b11} specifically studied the problem of learning meaningful assert statements and found it to be a significant challenge for automated approaches. Our study situates LLM-based generation in this landscape by asking whether contextual guidance resolves the fault-detection gap that coverage-oriented tools leave open.

\subsection{Human Testing Practices}

Bai et al.~\cite{b2} examined student and developer testing practices and found that human developers frequently prioritize common execution paths while neglecting edge cases and failure conditions, associating good testing primarily with high coverage even when coverage does not predict fault detection. These observations contextualize our finding that human-written tests in the BugsInPy dataset tend to target expected behavior rather than fault-inducing corner cases.

\subsection{Retrieval-Augmented Generation}

Lewis et al.~\cite{b18} introduced retrieval-augmented generation as a mechanism for supplying external documents to LLMs at inference time, improving factual accuracy and reducing hallucination. Shin et al.~\cite{b6} investigated RAG specifically for test generation and found that even lightweight retrieval improves generation relevance. Our implementation builds on these ideas but uses lexical overlap rather than embedding-based retrieval, reducing infrastructure requirements while retaining practical benefit. A key finding from our experiments is consistent with the RAG literature: excessive context dilutes prompt focus and degrades output quality.

\subsection{Mining Software Repositories for Benchmarks}

BugsInPy~\cite{b16} provides a curated collection of reproducible Python bugs drawn from real open-source projects. Studies using similar databases, including Defects4J~\cite{b15} for Java, have established them as standard evaluation vehicles for automated repair, test generation, and fault localization research. Zheng et al.~\cite{b14} argue more broadly that LLM code evaluation must move beyond correctness to consider multiple quality dimensions, a motivation that directly informs our metric design. Code-quality studies~\cite{b12,b13} have further shown that LLM-generated code often introduces maintainability concerns orthogonal to functional correctness, reinforcing the need for multi-dimensional evaluation.

\section{Study Design and Implementation}

\subsection{Overall Design}

We evaluate LLM-generated and human-written unit tests across three complementary benchmark settings. BugsInPy contributes ecological validity through reproducible historical faults; the open-source benchmark enables cleaner function-level comparison with less repository noise; and the controlled benchmark provides maximum experimental symmetry. No single benchmark is sufficient to characterize test quality in isolation, so the multi-dataset design is deliberate: it ensures that findings are not artifacts of any one dataset's characteristics.

The study pipeline proceeds in three stages: (1) structured benchmark construction, in which each bug or function is converted into a reproducible task package; (2) LLM test generation using a RAG-augmented pipeline; and (3) comparative evaluation across eight quality metrics.

\subsection{Benchmark Construction}

\subsubsection{BugsInPy Benchmark}

We used BugsInPy~\cite{b16} as the primary source of historical Python bugs. Starting from the full database, we applied three inclusion criteria: the bug must be reproducible in a Python~3 environment, at least one failing test must be present in the repository before the fix, and the changed source lines must be identifiable through a clean unified diff. After filtering, \textbf{29 bugs} were retained as the frozen benchmark set. Table~\ref{tab:benchmark} summarizes the benchmark composition.

For each selected bug we checked out both the buggy and fixed repository versions, extracted the unified diff, identified the functions and classes containing modified lines (patch context), and located all test files associated with the failing behavior. These artifacts were packaged into per-bug context folders consumed by both the generation and evaluation stages. Establishing a frozen, versioned benchmark set was essential: it ensures all subsequent stages operate on a stable, shared collection of tasks rather than a changing pool of candidates.

\subsubsection{Open-Source Function Benchmark}

To complement the historical bug setting we constructed a function-level benchmark from \texttt{python-slugify} and \texttt{packaging}. These projects were selected because they contain compact, well-documented utility functions with existing developer-written tests that serve as high-quality baselines. Each task consists of the target function, its module-level imports, and the corresponding human test file. Human-written tests were treated as locked baselines: the LLM received only the function source and the generation prompt, with no access to the human test during generation.

\subsubsection{Controlled Paired Benchmark}

The controlled benchmark consists of a smaller set of functions for which a researcher independently wrote reference tests under the same information conditions as the LLM (function source only, no bug context). This setting isolates the effect of the generation approach from differences in available context.

\begin{table*}[t]
\caption{Benchmark Composition}
\begin{center}
\resizebox{0.9\textwidth}{!}{%
\begin{tabular}{|l|l|c|l|}
\hline
\textbf{Benchmark} & \textbf{Source} & \textbf{\#Tasks} & \textbf{Human Baseline} \\
\hline
BugsInPy & Historical bugs from real open-source Python projects & 29 & Repository tests written before or independently of the specific fault \\
\hline
Open-source & Functions from \texttt{python-slugify} and \texttt{packaging} & 15 & Developer-written tests serving as locked baselines \\
\hline
Controlled & Researcher-designed functions with symmetric information conditions & 8 & Researcher-written tests with function source only (no bug context) \\
\hline
\textbf{Total} & & \textbf{52} & \\
\hline
\end{tabular}%
}
\end{center}
\label{tab:benchmark}
\end{table*}

\subsection{LLM Test Generation Pipeline}

We used Gemini~2.5~Flash~\cite{b20} as the generation model, selected for its balance of code-generation quality, inference speed, and API accessibility for iterative experimentation across many benchmark tasks.

Rather than relying on direct prompting alone, the pipeline integrates a lightweight lexical RAG mechanism. The context store for each task holds the bug patch diff, the surrounding source context, and relevant test-side artifacts. At generation time the system proceeds as follows.

\begin{enumerate}
    \item The target function's source is tokenized into a keyword set $K$.
    \item Each document $d$ in the context store is scored by normalized keyword overlap: $\text{score}(d) = |K \cap \text{tokens}(d)| \,/\, |K|$.
    \item The top $k{=}3$ documents are selected and appended to the prompt as reference material.
    \item The final prompt includes a system role (``expert Python test engineer''), the target function, the retrieved contexts, and explicit generation constraints: produce exactly one pytest test function, target only the provided source, and avoid extraneous helper code.
    \item The LLM response is parsed and the code block is extracted.
\end{enumerate}

Pilot experiments showed that $k > 3$ caused prompt dilution, increasing verbosity and reducing output consistency. The constrained prompt structure emerged from iterative refinement: initial broad prompts regularly produced verbose, off-target outputs with hallucinated helper functions and boilerplate unrelated to the target function.

\subsection{Evaluation Metrics}

Table~\ref{tab:metrics} defines the eight metrics used for comparison. \textbf{Fault detection} is the primary metric: a test is credited with fault detection if it fails on the buggy version and passes on the fixed version of the repository, acting as a regression test for the known defect. This operationalization follows standard practice in empirical testing studies~\cite{b17,b7}. Coverage is measured using \texttt{pytest-cov} at line and branch granularity. Assertion count, lines of code, and edge-case count are computed by static analysis of the test source. Documentation presence (docstring) and testing-pattern classification are likewise determined statically.

\begin{table}[htbp]
\caption{Evaluation Metrics}
\begin{center}
\scriptsize
\begin{tabular}{|p{1.5cm}|p{2.6cm}|p{2.4cm}|}
\hline
\textbf{Metric} & \textbf{What It Measures} & \textbf{Operationalization} \\
\hline
Fault Detection & Does the test expose the known bug? & Fails on buggy version, passes on fixed version \\
\hline
Line Coverage & \% of source lines executed & \texttt{pytest --cov} \\
\hline
Branch Coverage & \% of decision paths taken & \texttt{pytest --cov-branch} \\
\hline
Assertion Count & Verification thoroughness & Count \texttt{assert} statements \\
\hline
Lines of Code & Test verbosity & Count non-blank, non-comment lines \\
\hline
Edge Cases & Input-space diversity & Count distinct input values per test \\
\hline
Documentation & Self-documentation quality & Docstring present (binary) \\
\hline
Testing Patterns & Approach sophistication & Simple / parametrized / mock / advanced \\
\hline
\end{tabular}
\end{center}
\label{tab:metrics}
\end{table}

\subsection{Statistical Analysis}

Binary outcomes (fault detection, docstring presence) are compared using Fisher's exact test with Cohen's~$h$ as the effect size. Continuous and ordinal metrics (lines of code, assertion count, edge-case count, coverage percentages) are compared using the two-sided Mann-Whitney $U$ test with rank-biserial correlation $r$ as the effect size. All tests use a significance threshold of $\alpha = 0.05$. Effect sizes follow Cohen's conventions: $|h|$ or $|r| \geq 0.5$ indicates a large effect.

\section{Results}

\subsection{RQ1: Fault Detection}

\begin{figure}[H]
\centerline{\includegraphics[width=0.48\textwidth]{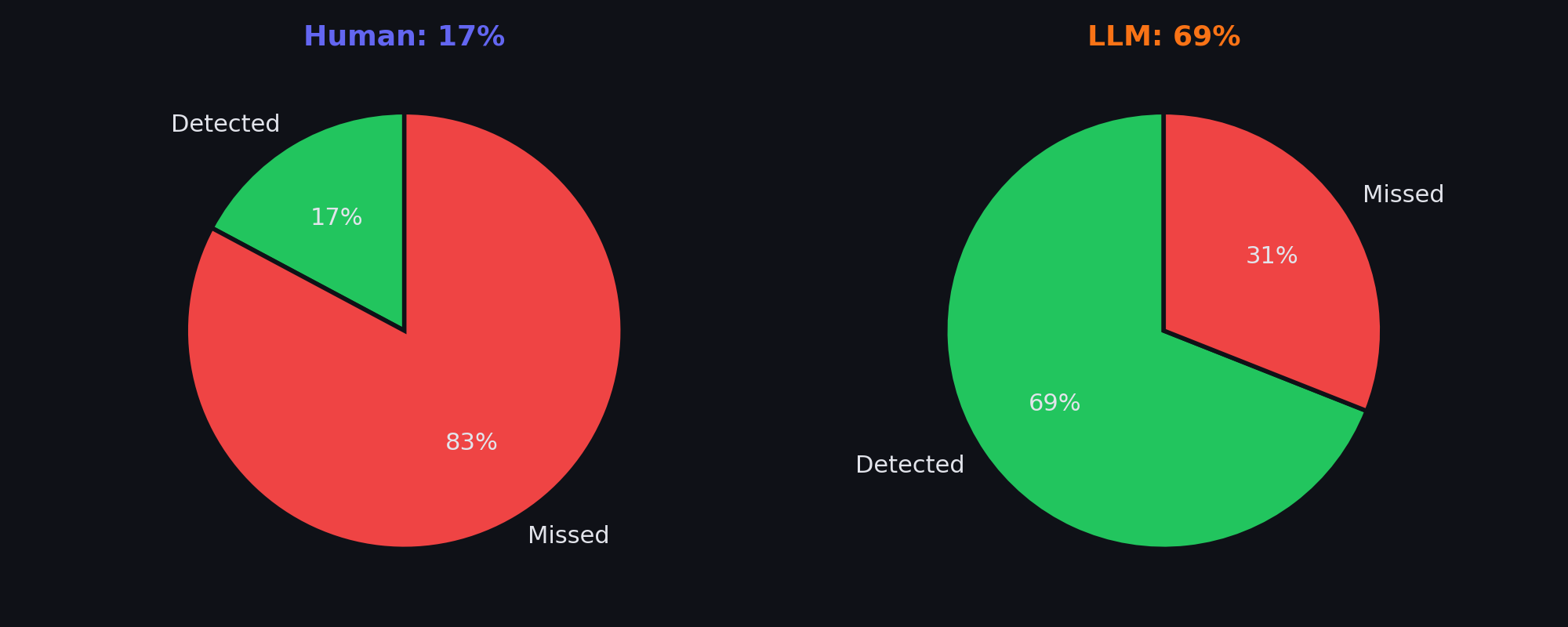}}
\caption{Fault detection comparison on 29 BugsInPy bugs. LLM tests with RAG context detected 20 of 29 faults (69\%); human-written general-purpose tests detected 5 of 29 (17.2\%). Fisher's exact: $p < 0.001$, Cohen's~$h = 1.10$.}
\label{fig:fault_detection}
\end{figure}

LLM-generated tests detected faults in 20 of 29 cases (69.0\%) compared to 5 of 29 (17.2\%) for human-written tests (Figure~\ref{fig:fault_detection}). Fisher's exact test confirms the difference is highly significant ($p < 0.001$, Cohen's $h = 1.10$, large effect).

\textbf{Interpreting the comparison.} This result requires careful framing. LLM tests are generated using retrieved bug context, including the patch diff and bug description, making them specifically targeted regression tests. Human-written tests in the BugsInPy repositories were written before or independently of the specific fault and represent general-purpose assertions. The comparison therefore answers: when bug context is available, can an LLM generate a regression test that catches the known fault more reliably than a pre-existing general test? The answer is clearly yes. The practical implication, discussed further in Section~\ref{sec:discussion}, is that LLM generation is most valuable at fix time, when this context naturally exists.

\subsection{RQ2: Conditions for LLM Advantage or Disadvantage}

Our analysis identifies four conditions under which LLM tests consistently outperform human baselines and three conditions under which human tests retain a practical advantage.

\textbf{LLM-generated tests are stronger when:}
\begin{itemize}
    \item \textit{Bug context is available.} The RAG pipeline is essential to the observed fault-detection advantage. Organizations seeking to adopt LLM test generation should build retrieval infrastructure around their defect-tracking systems and patch review workflows.
    \item \textit{Exhaustive input coverage is required.} LLMs readily emit parametrized tests covering broad input ranges, a pattern humans use sparingly (24\% vs.\ 10\% of tests, respectively).
    \item \textit{Comprehensive documentation is valued.} LLM tests include docstrings in 65.5\% of cases versus 0\% for human-written tests in this benchmark (Fisher's exact, $p < 0.001$, Cohen's~$h = 1.28$).
    \item \textit{Edge-case diversity is the goal.} LLMs produced an average of 5.0 distinct input scenarios per test compared to 3.0 for human tests.
\end{itemize}

\textbf{Human-written tests are stronger when:}
\begin{itemize}
    \item \textit{No prior fault knowledge is available.} Without bug context, LLM tests are unlikely to target fault-specific code paths, and the RAG advantage disappears.
    \item \textit{Conciseness and long-term maintainability are priorities.} Human tests average 9.6 lines of code versus 31.0 for LLM tests, making them substantially easier to read and modify in active codebases (Mann-Whitney $U$, $p < 0.001$, large effect).
    \item \textit{Deep domain or behavioral knowledge is required.} Human developers encode implicit understanding of acceptable behavior that may not be recoverable from function source alone.
\end{itemize}

\subsection{RQ3: Coverage, Assertions, and Testing Patterns}

\subsubsection{Coverage Analysis}

Table~\ref{tab:results} reports quality and coverage metrics across all 52 tasks. The most striking finding is coverage parity: human and LLM tests achieve nearly identical line coverage (84.8\% vs.\ 88.5\%, Mann-Whitney $p = 0.28$) and branch coverage (75.2\% vs.\ 82.1\%, Mann-Whitney $p = 0.17$). These differences are not statistically significant. Despite this similarity, fault-detection rates differ by a factor of four (17.2\% vs.\ 69\%). This result provides direct empirical support for the position that coverage is an insufficient proxy for test quality~\cite{b17}: a test suite may execute nearly all code paths while still missing the specific assertion that would fail on faulty behavior.

\begin{table}[H]
\caption{Quality and Coverage Results (means across 52 tasks; fault detection on 29 BugsInPy bugs)}
\begin{center}
\begin{tabular}{|l|c|c|c|}
\hline
\textbf{Metric} & \textbf{Human} & \textbf{LLM} & \textbf{Sig.} \\
\hline
Fault Detection & 17.2\% & 69.0\% & $p < 0.001$ \\
\hline
Line Coverage & 84.8\% & 88.5\% & $p = 0.28$ \\
\hline
Branch Coverage & 75.2\% & 82.1\% & $p = 0.17$ \\
\hline
Assertions (avg.) & 3.2 & 5.5 & $p < 0.01$ \\
\hline
Lines of Code (avg.) & 9.6 & 31.0 & $p < 0.001$ \\
\hline
Edge Cases (avg.) & 3.0 & 5.0 & $p < 0.01$ \\
\hline
Docstring (\%) & 0\% & 65.5\% & $p < 0.001$ \\
\hline
\end{tabular}
\end{center}
\label{tab:results}
\end{table}

\begin{figure}[H]
\centerline{\includegraphics[width=0.4\textwidth]{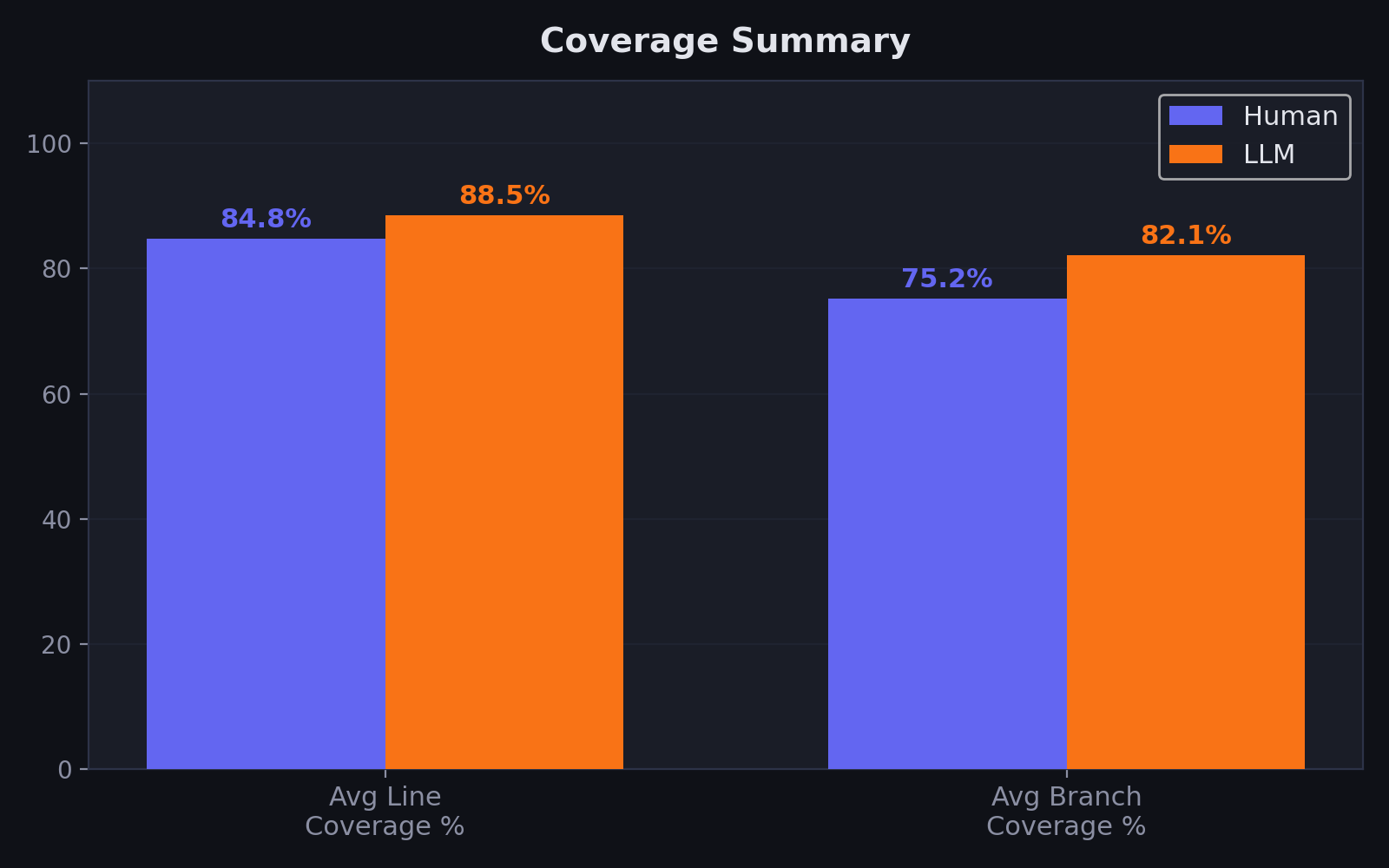}}
\caption{Coverage comparison across all 52 tasks. Line and branch coverage differences are not statistically significant ($p > 0.15$ for both), despite a four-fold difference in fault-detection rate.}
\label{fig:coverage}
\end{figure}

\subsubsection{Verbosity and Assertions}

LLM-generated tests are substantially more verbose: 31.0 lines on average versus 9.6 for human tests, a 3.2$\times$ difference significant at $p < 0.001$ with a large effect. Assertion counts also differ significantly (5.5 vs.\ 3.2, $p < 0.01$), as does edge-case coverage (5.0 vs.\ 3.0, $p < 0.01$). While higher assertion density and broader input coverage are generally desirable properties, the verbosity increase raises legitimate concerns about maintainability in codebases where concise, readable tests are preferred.

\subsubsection{Testing Patterns and Quality Profile}

\begin{figure}[H]
\centerline{\includegraphics[width=0.48\textwidth]{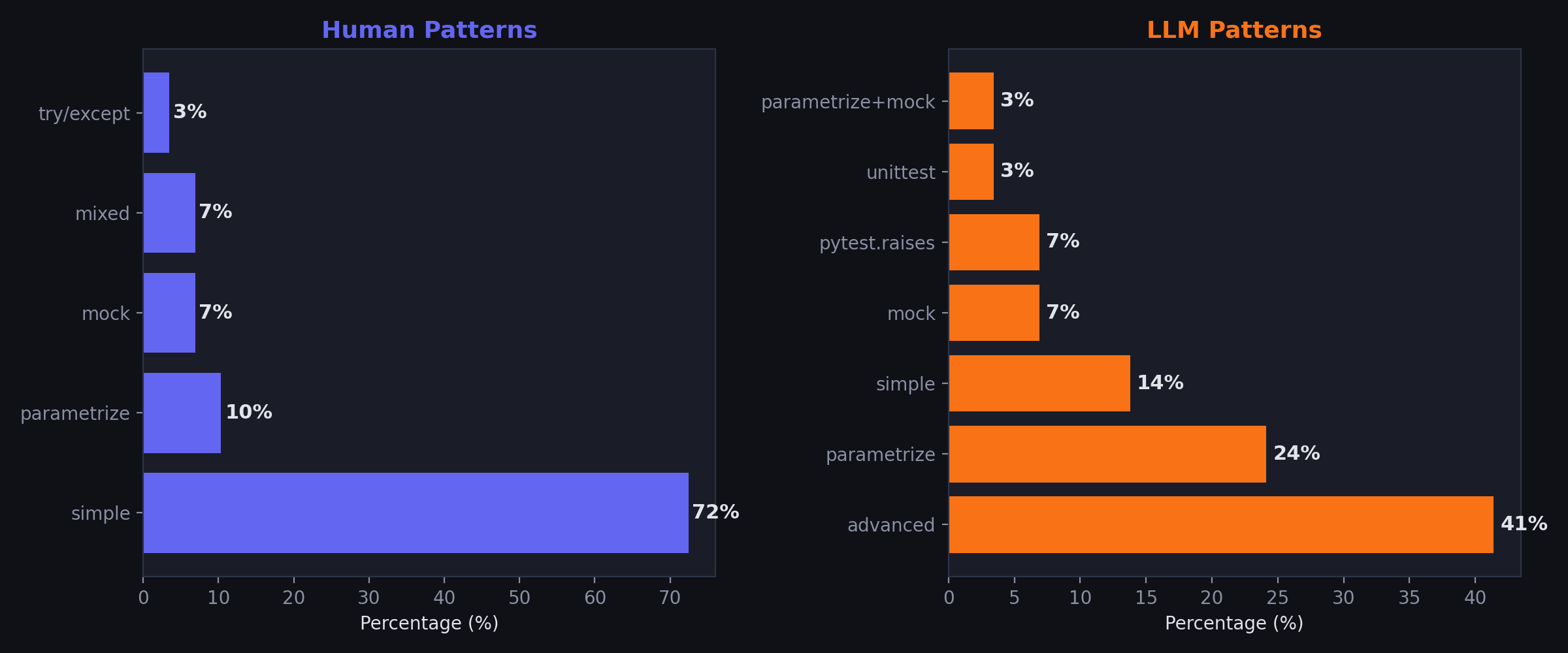}}
\caption{Testing-pattern distribution. Humans favor simple assertions (72\%); LLMs use advanced composite patterns (41\%) and parametrized testing (24\%) far more frequently.}
\label{fig:patterns}
\end{figure}

Figure~\ref{fig:patterns} shows the distribution of testing patterns. Human-written tests rely predominantly on simple assertions (72\%), with modest use of parametrized testing (10\%), mock objects (7\%), and try/except blocks (7\%). LLM-generated tests exhibit a markedly different profile: advanced composite patterns dominate (41\%), followed by parametrized testing (24\%), simple assertions (14\%), \texttt{pytest.raises} exception testing (7\%), and mock objects (7\%).

\begin{figure}[H]
\centerline{\includegraphics[width=0.38\textwidth]{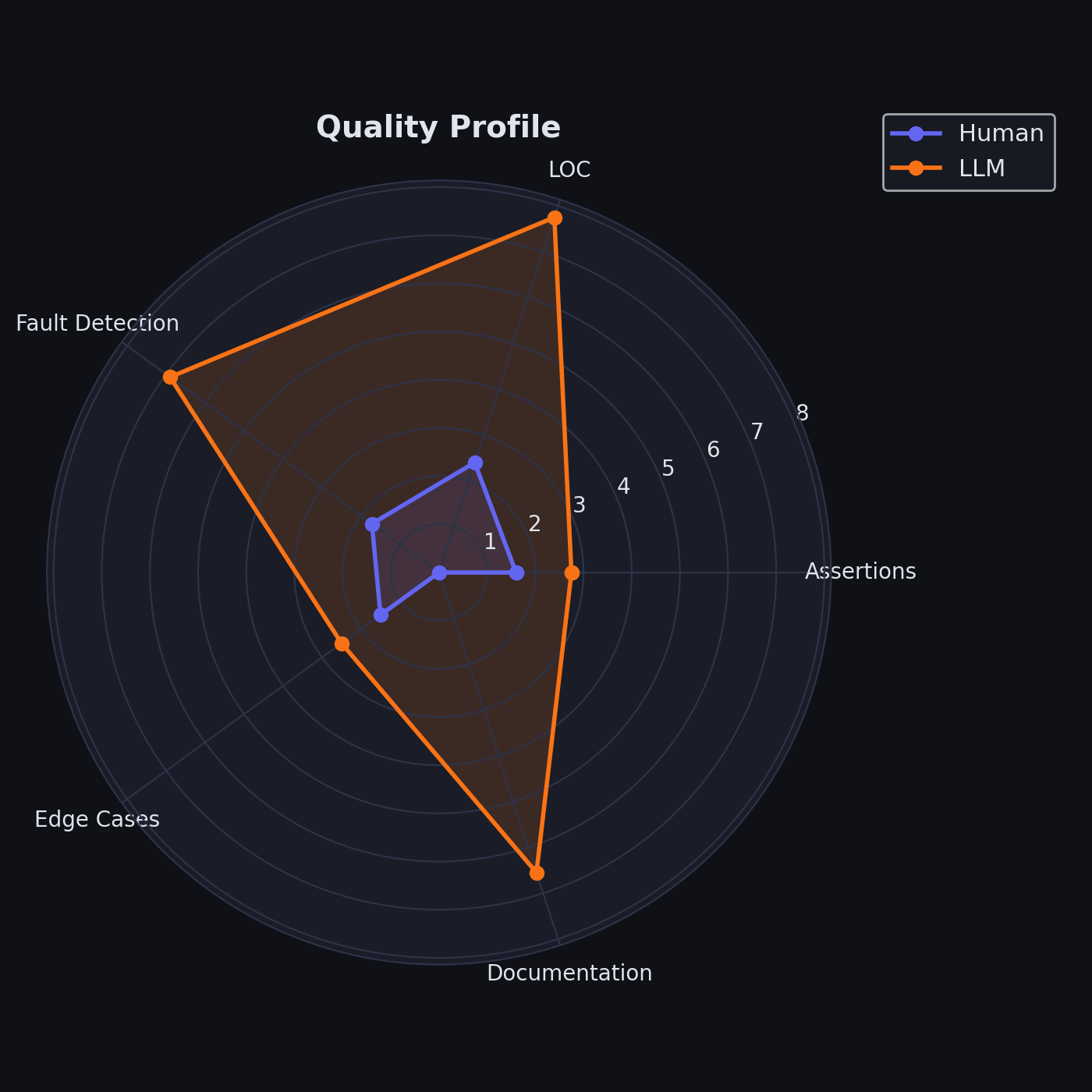}}
\caption{Quality profile radar comparing human and LLM tests across five dimensions. The LLM profile is larger on most axes; the human profile is more compact, reflecting an emphasis on conciseness.}
\label{fig:quality_radar}
\end{figure}

The quality profile radar in Figure~\ref{fig:quality_radar} summarizes these differences visually. LLM tests score higher on most dimensions, but human tests achieve their fault-detection rate with significantly less code, suggesting higher efficiency in terms of defects caught per line of test code.

\begin{figure}[H]
\centerline{\includegraphics[width=0.48\textwidth]{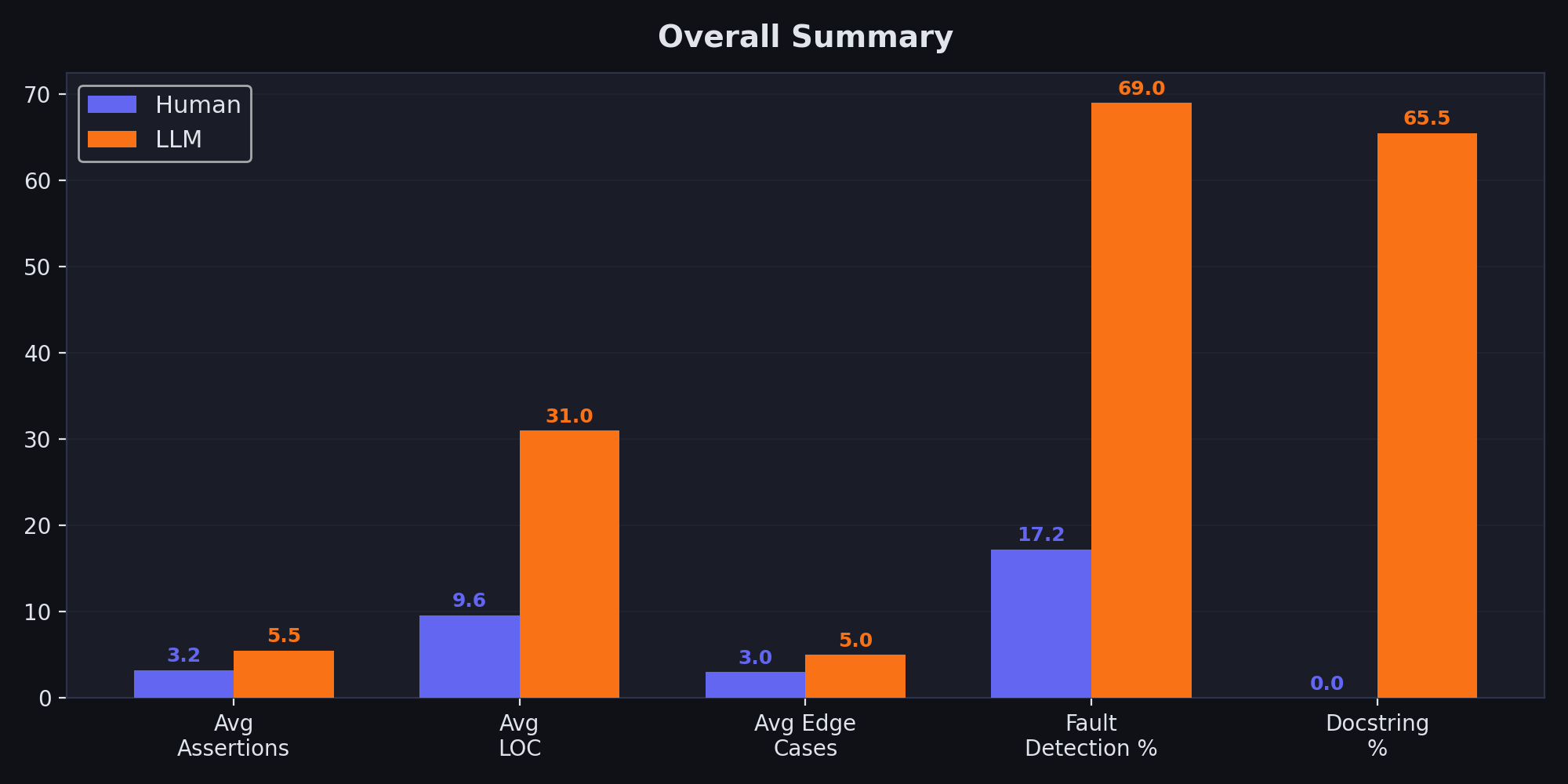}}
\caption{Summary comparison across five metrics. LLM tests score higher on assertions, lines of code, edge cases, fault detection, and docstring rate; human tests are more concise.}
\label{fig:overall_summary}
\end{figure}

\section{Discussion}
\label{sec:discussion}

\subsection{Regression Testing as the Natural LLM Use Case}

The fault-detection gap should not be read as evidence that LLMs are simply better test writers in all contexts. It reflects a specific and practically important scenario: when a defect has been identified and its patch is under review, an LLM with access to that context can generate a targeted regression test far more reliably than a general-purpose test written without foreknowledge of the defect. This framing has direct implications for how teams should deploy LLM test generation. The highest-value integration point is at fix time, as part of the patch review workflow: generate an LLM test alongside the fix, provide it with the diff and bug report, and use it as an automatic regression guard.

Human-written tests remain indispensable for general-purpose coverage and for capturing behavioral expectations that are not apparent from function source alone. Our data show that coverage levels are nearly identical between approaches, which means LLM regression tests do not substantially expand the code paths exercised beyond what human tests already cover. The two approaches are complementary: human tests provide broad behavioral validation; LLM regression tests provide targeted fault-detection at defect-resolution time.

\subsection{Qualitative Example: \texttt{parse\_dfxp\_time\_expr}}

To ground the aggregate statistics in a concrete case, we examine tests generated for \texttt{parse\_dfxp\_time\_expr()} from \texttt{youtube-dl}, a function that parses DFXP subtitle time expressions and converts valid strings to floating-point second values.

\begin{figure}[H]
\centerline{\includegraphics[width=0.42\textwidth]{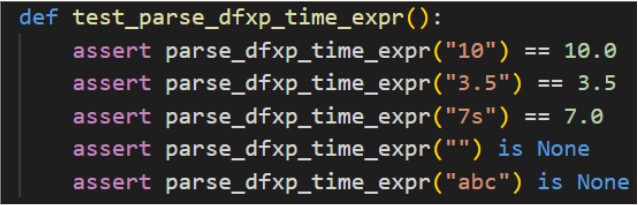}}
\caption{Human-written test for \texttt{parse\_dfxp\_time\_expr()}, focusing on common valid and invalid inputs.}
\label{fig:human_dfxp}
\end{figure}

\begin{figure}[H]
\centerline{\includegraphics[width=0.42\textwidth]{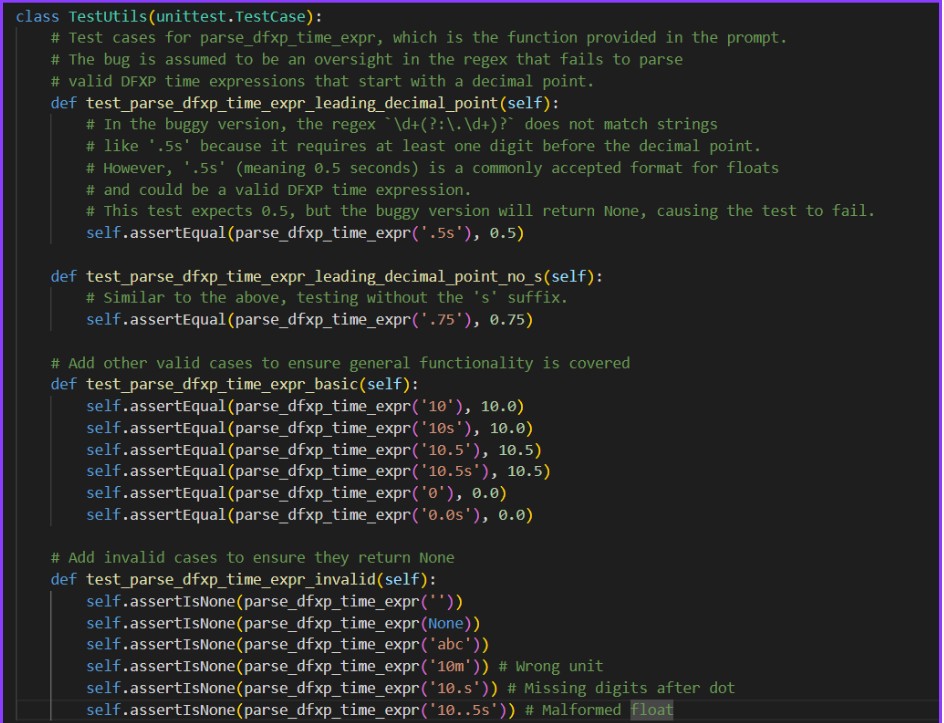}}
\caption{LLM-generated test for \texttt{parse\_dfxp\_time\_expr()}, covering edge cases and malformed inputs not addressed by the human test.}
\label{fig:llm_dfxp}
\end{figure}

The human-written test (Figure~\ref{fig:human_dfxp}) validates common integer and floating-point inputs, standard time suffix handling, and a small number of explicitly invalid cases. The LLM-generated test (Figure~\ref{fig:llm_dfxp}) covers a wider range of malformed inputs, including leading-decimal formats (\texttt{.5s}) and double-decimal expressions (\texttt{10..5s}) that the original regular expression does not handle. Several of these inputs reveal silent failures in the parser, precisely the kind of fault-detection gap visible in the aggregate statistics.

This example also illustrates the verbosity trade-off. The LLM test contains additional assertions and a docstring that improve self-documentation, but the test is over three times longer than the human version and includes boilerplate that a developer would likely trim for production use.

\subsection{Practical Recommendations}

Three recommendations follow from the findings.

\begin{enumerate}
    \item \textbf{Deploy LLM generation at fix time.} When a defect is confirmed and a patch is under review, generate LLM tests with the bug diff and description attached. The retrieved context is sufficient to produce high-precision regression tests without embedding or vector-database infrastructure.
    \item \textbf{Do not replace human test suites with LLM output.} Human tests encode behavioral knowledge that LLMs cannot infer from source alone. Coverage parity between approaches shows that LLM tests do not substantially expand behavioral coverage beyond existing human suites.
    \item \textbf{Evaluate tests on fault detection rather than coverage alone.} Our data confirm that coverage masks large differences in fault-detection capability. Incorporating mutation testing or regression test effectiveness into continuous integration quality gates provides a substantially stronger signal than line or branch coverage.
\end{enumerate}

\section{Threats to Validity}

\subsection{Internal Validity}

The most significant internal threat is the information asymmetry between LLM and human tests. LLM tests receive bug patch diffs and descriptions via the RAG pipeline; human-written tests were developed without foreknowledge of the specific fault. This asymmetry means the fault-detection comparison reflects the value of having bug context at generation time, not an unconfounded head-to-head comparison of generation capability. We mitigate this by framing RQ1 explicitly as a comparison of context-aware regression generation against general-purpose testing, and by noting in Section~\ref{sec:discussion} that a direct capability comparison would require human testers given identical contextual information.

A secondary internal threat is test-collection inconsistency. Different benchmark components were prepared by different team members using slightly varying extraction procedures, potentially introducing formatting or context-depth variation. We mitigated this through a normalization pass and a shared annotation schema, but residual inconsistency cannot be fully ruled out.

\subsection{External Validity}

The BugsInPy component covers 29 bugs from a limited set of Python open-source projects. Findings may not generalize to other programming languages, proprietary codebases, or domains with substantially different defect distributions. The open-source benchmark is limited to two small utility libraries, restricting functional diversity. Extending to larger and more varied repositories, and to Java via Defects4J~\cite{b15}, would strengthen generalizability.

All results are based on a single LLM (Gemini~2.5~Flash). Different models, especially those with stronger formal reasoning or code understanding, may produce materially different results.

\subsection{Construct Validity}

Fault detection is operationalized as a test that fails on the buggy version and passes on the fixed version. This is standard and well-motivated~\cite{b16,b17} but does not capture tests that correctly describe expected behavior for a different aspect of the function without triggering the specific mutation in the patch. Edge-case count and testing-pattern classification are computed by static analysis, which may misclassify atypically structured tests. Docstring presence is a weak proxy for documentation quality; a one-line docstring counts identically to a comprehensive one.

\subsection{Conclusion Validity}

With 29 bugs in the fault-detection analysis, the study has limited statistical power for subgroup comparisons (e.g., per-project or per-pattern breakdowns). We use Fisher's exact tests and Mann-Whitney $U$ tests throughout because they are appropriate for small samples without normality assumptions. Subgroup analyses should nonetheless be treated as exploratory rather than confirmatory.

\section{Conclusion}

We presented an empirical comparison of LLM-generated and human-written unit tests across three complementary Python benchmarks. Using Gemini~2.5~Flash augmented with a lightweight lexical RAG pipeline, we found that context-aware LLM tests detect faults in 69\% of 29 real historical bugs compared to 17.2\% for general-purpose human-written tests (Fisher's exact, $p < 0.001$, Cohen's~$h = 1.10$). The advantage is not attributable to higher coverage: line and branch coverage are statistically indistinguishable between the two approaches, directly demonstrating that coverage is an insufficient proxy for fault-detection capability.

The central practical insight is that LLM test generation is most valuable in the regression testing context, where bug context is available to the generation pipeline. In this setting, LLM tests provide substantially stronger fault-detection guarantees than pre-existing general tests. For general-purpose testing without prior fault context, the advantage diminishes; human tests remain more concise, more maintainable, and comparably effective in terms of code coverage. The two approaches are therefore complementary: human tests for ongoing behavioral validation, LLM tests for targeted regression coverage at fix time.

Avenues for future work include comparing multiple LLMs across the same benchmark suite, extending to Java and Defects4J~\cite{b15}, conducting a controlled experiment in which human developers receive identical bug context to the LLM, and investigating hybrid workflows that combine human and LLM tests to capture the strengths of both approaches.

\end{document}